\documentclass[12pt]{article}
\usepackage{amssymb}

\input{tcilatex}

\begin{document}

\bigskip \baselineskip1cm \textwidth16.5 cm

\bigskip

\begin{center}
\textbf{FINITE-SIZE SCALING AND POWER LAW RELATIONS FOR DIPOL-QUADRUPOL
INTERACTION ON BLUME-EMERY-GRIFFITHS MODEL }

\textbf{\ }

Aycan \"{O}zkan, B\"{u}lent Kutlu

Gazi \"{U}niversitesi, Fen -Edebiyat Fak\"{u}ltesi, Fizik B\"{o}l\"{u}m\"{u}%
, 06500 Teknikokullar, Ankara, Turkey,

aycan@gazi.edu.tr, bkutlu@gazi.edu.tr
\end{center}

\textbf{Abstract:} The Blume-Emery-Griffiths model with the dipol-quadrupol
interaction ($\ell $ ) has been simulated using a cellular automaton
algorithm improved from the Creutz cellular automaton (CCA) on the face
centered cubic (fcc) lattice. The finite-size scaling relations and the
power laws of the order parameter ($M$) and the susceptibility ($\chi $) are
proposed for the dipol-quadrupol interaction ($\ell $ ). The dipol-quadrupol
critical exponent $\delta _{\ell }$ has been estimated from the data of the
order parameter ($M$) and the susceptibility ($\chi $). The simulations have
been done in the interval $0\leq \ell =L/J\leq 0.01$ for $d=D/J=0$, $k=K/J=0$
and $h=H/J=0$ parameter values on a face centered cubic (fcc) lattice with
periodic boundary conditions. The results indicates that the effect of the $%
\ell $ parameter is similar to the external magnetic field ($h$). The
critical exponent $\delta _{\ell }$ are in good agreement with the universal
value ($\delta _{h}=5$) of the external magnetic field.

\textbf{Keywords:} Cellular automaton, face-centered cubic,
Blume-Emery-Griffiths model, critical exponent

The spin-1 Ising model, which is known as the generalized
Blume-Emery-Griffiths (BEG) model, can be used to simulate many physical
systems. The model firstly has been presented for describing phase
separation and superfluid ordering in He mixtures $\left[ 1\right] $. The
most general Hamiltonian of the model is given by \qquad

\begin{equation}
H_{I}=-J\sum_{<ij>}S_{i}S_{j}-K\sum_{<ij>}S_{i}^{2}S_{j}^{2}+L%
\sum_{<ij>}(S_{i}^{2}S_{j}+S_{i}S_{j}^{2})+D\sum_{i}S_{i}^{2}+H\sum_{i}S_{i}
\end{equation}%
which is equivalent to the lattice gas Hamiltonian under some
transformations $\left[ 2\right] -\left[ 4\right] $. $\left\langle
ij\right\rangle $ denotes summation over all nearest-neighbor (nn) pairs of
sites and $s_{i}=-1,$ $0,$ $1$. The parameters $J$, $K$, $L$, $D$ and $H$
are bilinear, biquadratic, dipole-quadrupole interaction terms, the
single-ion anisotropy constant and the external field term. The versions of
the model have been applied to the physical systems such as the
solid-liquid-gas systems $\left[ 5\right] $, the multicomponent fluids $%
\left[ 6\right] $, the microemulsions $\left[ 7\right] $, the semiconductor
alloys $\left[ 8\right] -\left[ 10\right] $, He$^{3}$-He$^{4}$ mixtures $%
\left[ 1,11\right] $,the binary alloys $\left[ 12\right] $ and the magnetic
spin systems $\left[ 13\right] -\left[ 16\right] $.

The spin-1 Ising model with dipole-quadrupole interaction was firstly
studied using molecular field approximation $\left[ 6,17,18\right] $ and the
transfer matrix method $\left[ 19\right] $ to investigate the tricritical
and multicritical points for selected values of the $\ell $ parameter on one
and two dimension. The spin-1 Ising model has also applied to two
dimensional\ ternary graphite intercalation compounds (GIC's) for
investigating the influence of the dipole-quadrupole interaction on the
melting point using Monte Carlo method $\left[ 20\right] $. The model has
been used for estimating the critical concentration value of the (GaAs)$%
_{1-x}$Ge$_{2x}$ alloy by Kikuchi approximation $\left[ 9,10\right] $ and
Cluster variation method $\left[ 21,22\right] $. These studies show that the
dipol-quadrupol interaction term is very effective on the critical
concentration value of the ($III-V$)$_{1-x}IV_{x}$ ternary alloys $\left[ 23%
\right] $ and it can be dominant on the phase space $\left[ 18,19,24\right] $%
. However, the infinite lattice critical behaviors of the order parameter ($M
$) and the susceptibility ($\chi $) with the dipol-quadrupol interaction ($%
\ell =L/J$ ) has not been studied yet. In our previous paper $\left[ 25%
\right] $, it was shown that the dipol-quadrupol interaction ($\ell $)
prevents the one of the spin species ($S=+1$ or $-1$) on the BEG model ($%
J\neq 0$, $K\neq 0$, $D\neq 0$) and breaks the symmetry in the Hamiltonian
similar to the external magnetic field ($h$) $\left[ 26\right] $. As a
result of this, the dense ferromagnetic ($df$($+$), $df$($-$)) and
ferromagnetic ($F$($+$), $F$($-$)) phases occur on the ground state phase
diagram for nonzero $\ell $\ value. Furthermore, model exhibits the
reentrant, successive and multi phase transitions in the interval $-0.15\leq
\ell \leq 0.15$ and it does not show any phase transition at the large $\ell 
$ value ($\ell >0.15$).

The aim of this study is to investigate the critical behavior of the BEG
model with the dipol-quadrupol interaction ($\ell $ ) and to estimate the
power laws and the finite size scaling relations for the order parameter ($M$%
) and the susceptibility ($\chi $) in the $\ell =L/J>0$ region. In order to
expose the effects of the dipol-quadrupol interaction ($\ell =L/J$), the
biquadratic interaction ($k=K/J$), the single-ion anisotropy term and the
external magnetic field ($h=H/J$) have been taken zero ( $d=0$, $k=0$ and $%
h=0$).\ These terms were studied in our previous papers and the effects of
them on the phase space and the critical behavior of the BEG model were
discussed $\left[ 25\right] -\left[ 29\right] $.

The temperature dependence of the order parameters ($M$, $Q$), the
susceptibility ($\chi $), the specific heat ($C/k$) and the Ising energy ($%
H_{I}$) have been computed on the fcc lattice with linear dimension L$=4$, $%
6 $, $8$, $9$ and $12$. The finite lattice critical temperatures are
estimated from the maxima of the susceptibility ($\chi $) for the fcc
lattice with periodic boundary conditions. In order to expose the influence
of the dipole-quadrupole interaction, the thermodynamic quantities have been
calculated using CA heating $\left[ 25\right] $, $\left[ 27\right] -$ $\left[
29\right] $ algorithm on the fcc lattice for L$=4$, $6$, $8$, $9$ and $12$
(The total number of sites is $N=4$L$^{3}$) with the periodic boundary
conditions. The fcc lattice can be built from four interpenetrating simple
cubic lattices. The linear dimension L$=12$ of face centered cubic lattice
corresponds to the L$=20$ in simple cubic lattice. Hence, the data are
analyzed within the framework of the finite-size scaling theory for the BEG
model with the dipol-quadrupol interaction ($\ell $ ).

The Creutz cellular automaton (CCA) is faster than the conventional Monte
Carlo method (MC) $\left[ 31\right] $. The CCA does not need high quality
random numbers and it is a new and an alternative simulation method for
physical systems $\left[ 25\right] -\left[ 29\right] $, $\left[ 30\right] -%
\left[ 33\right] $. Furthermore, the results obtained using CA algorithm and
its improved versions are in good agreement with the universal critical
behavior for the BEG model. Thus the simulations have been carried out using
a cellular automaton heating algorithm which successfully produces the
critical behavior of the Ising model $\left[ 25\right] $, $\left[ 27\right] -%
\left[ 29\right] $. During the heating cycle, energy is added to the spin
system through the second variables ($H_{K}$) after the $2.000.000$ cellular
automaton steps. The heating rate is equal to $0.08H_{K}$ per site. The
computed values of the thermodynamic quantities are averages over the
lattice and the number of time steps ($2.000.000$ ) with discard of the
first $100.000$ time steps during the cellular automaton develops. For the
finite size lattice, the order parameters $M$ and $Q$ are given by

\begin{equation}
M=\frac{1}{\text{L}^{3}}\sum_{i}S_{i}
\end{equation}

\begin{equation}
Q=\frac{1}{\text{L}^{3}}\sum_{i}S_{i}^{2}
\end{equation}%
The susceptibility and the specific heat are calculated with

\begin{equation}
\chi =\text{L}^{3}\frac{\left\langle M^{2}\right\rangle -\left\langle
M\right\rangle ^{2}}{kT}
\end{equation}

\begin{equation}
C=\text{L}^{3}\frac{\left\langle H_{I}^{2}\right\rangle -\left\langle
H_{I}\right\rangle ^{2}}{kT}
\end{equation}%
The expectation values in equation (5) and (6) are averages over the lattice
and the number of the time steps.

In order to investigate the critical behavior of the BEG model with the
dipole-quadrupole interaction, the thermodynamic quantities have been
calculated using CA heating algorithm on the fcc lattice for L$=4$, $6$, $8$%
, $9$ and $12$ (The total number of sites is $N=4$L$^{3}$) with the periodic
boundary conditions.

For the $d=0$ and the $k=0$ line, the model exhibits the second order
Ferromagnetic-Paramagnetic phase transition $\left[ 25\right] $ for $\ell =0$
while the susceptibility ($\chi $) have a characteristic peak. With the
increasing $\ell $ value in the interval $0\leq \ell =L/J\leq 0.01$, the
characteristic peak of susceptibility occurs at the higher critical
temperature and its peak value decreases. The effect of the $\ell $
parameter is similar to those coming from the external magnetic magnetic
field $\left[ 25,26,33\right] $.

At $T=T_{C}($L$)$, the order parameter values ( $M(T_{C}($L$),\ell )$ ) have
been estimated for the increasing $\ell $ value in the interval $0\leq \ell
=L/J\leq 0.01$ on the different lattice sizes L$=4$, $6$, $8$, $9$ and $12$
and illustrated in figure (1a). The results show that there is a
considerable finite size effect near the $\ell =0$. The $M(T_{C}($L$),\ell )$
tends to zero with the increasing lattice sizes at $\ell =0$. However the $%
M(T_{C}($L$),\ell )$ goes to the constant value ($\sim 0.5$) for all lattice
size with the increasing value of $\ell .$ The susceptibility data $\chi
(T_{C}($L$),\ell )$ \ are plotted in figure (1b)\textit{\ }for\textit{\ }$%
\ell $ values in the interval $0\leq \ell \leq 0.01$ on the different
lattice sizes L$=4$, $6$, $8$, $9$ and $12$\textit{. }Figure (1b) obviously
shows that\textit{\ }the susceptibility data $\chi (T_{C}($L$),\ell )$ tend
to infinity with increasing lattice size at $\ell =0.$ Around the $\ell =0$
value, the data show the dependence on the lattice size and $\chi (T_{C}($L$%
),\ell )$ is a decreasing function of \ $\ell $ for the all lattice size.

The dipol-quadrupol interaction on the spin system shows a similar effect
with the external magnetic field. As it is seen in figure (1), the order
parameter ($M$) and the susceptibility ($\chi $) are the functions of $\ell $
. For the determination of power-law exponent $\delta _{\ell }$, the power
law relations of the thermodynamic functions ( $M$ and $\chi $) for the
dipol-quadrupol interaction ($\ell $) have been considered as similar
expressions to the power-laws of external magnetic field ($h$) $\left[ 26,34%
\right] $. Therefore, the power laws of the order parameter ($M$) and the
susceptibility ($\chi $) at $T=T_{C}$ have been described by

\begin{equation}
M(T_{C},\ell )=B_{\ell }\left\vert \ell \right\vert ^{1/\delta _{\ell }}
\end{equation}

\begin{equation}
\chi (T_{C},\ell )=C_{\ell }\left\vert \ell \right\vert ^{(1-\delta _{\ell
})/\delta _{\ell }}
\end{equation}%
where $B_{\ell }$ and $C_{\ell }$ are the order parameter and the
susceptibility amplitudes. In the interval $0\leq \ell \leq 0.01$, the order
parameter ($M$), the critical exponent $\delta _{\ell }($L$)$\ and the order
parameter amplitude $B_{\ell }($L$)$ on each lattice size (L) are obtained
from the best fit to straight lines in the Log-Log plot of the data (Figure
(2a)). The estimated $\delta _{\ell }($L$)$ and $B_{\ell }($L$)$ values are
plotted against L$^{-1/\nu }$ in figure (2b) and (c). The extrapolations (L$%
^{-1/\nu }$ $\rightarrow 0$) of data which lie on straight lines give the
infinite lattice values as $1/\delta _{\ell }(\infty )=0.193\pm 0.005$ and $%
B_{\ell }(\infty )=0.98\pm 0.06$. The value $\delta _{\ell }=5.18$ is in
good agreement with universal value ($\delta _{h}=5$) for external magnetic
field $\left[ 26,34\right] $.

In figure 3, the critical exponent $\delta _{\ell }($L$)$\ and the
susceptibility amplitude $C_{\ell }($L$)$ are obtained from the
susceptibility data ($\chi (T_{C}($L$),\ell )$ ). The Log-Log plot of $\chi
(T_{C}($L$),\ell )$ against $\ell $ in the interval $0\leq \ell \leq 0.001$
yields $(1-\delta _{L}($L$))/\delta _{L}($L$)$ and $C_{\ell }($L$)$ in
figure (3a). The extrapolation of these quantities gives $(1-\delta _{\ell
}(\infty ))/\delta _{\ell }(\infty )=-0.80\pm 0.05$ and $C_{\ell }(\infty
)=0.008\pm 0.001$ (Figure (3b) and 3(c)). Therefore, the value of $\delta
_{\ell }(\infty )$ is obtained as $5$ which is equal to the universal value $%
\delta _{h}=5$ for the external magnetic field.

The finite-size scaling relations $[26]$ of the order parameter ($M$) and
the susceptibility ($\chi $) related to the dipol-quadrupol interaction $%
\ell $ can be defined by scaling relations related to the external magnetic
field $h$ as

\begin{equation}
M=\text{L}^{-\beta /\nu }X^{o}(\text{L}^{\delta _{\ell }\beta /\nu
}\left\vert \ell \right\vert ,\text{L}^{1/\nu }\varepsilon )
\end{equation}

\begin{equation}
kT\chi =L^{\gamma /\nu }Y^{o}(\text{L}^{\gamma \delta _{\ell }/\nu (\delta
_{\ell }-1)}\left\vert \ell \right\vert ,\text{L}^{1/\nu }\varepsilon )
\end{equation}

with $\varepsilon =(T-T_{C}(\infty ))/T_{C}(\infty )$ $\left[ 26,34\right] $.

At $T=T_{C}(\infty )$, the scaling functions $X^{\circ \text{ }}$and $%
Y^{\circ }$ are asymptotically reproduced as%
\begin{equation}
X^{\circ }(x)=B_{\ell }x^{1/\delta _{\ell }}
\end{equation}

\begin{equation}
Y^{\circ }(x)=C_{\ell }x^{(1-\delta _{\ell })/\delta _{\ell }}
\end{equation}%
\ where $x=$L$^{\delta _{\ell }\beta /\nu }\left\vert \ell \right\vert $, $%
B_{\ell }$ and $C_{\ell }$ are the order parameter and the susceptibility
amplitudes.\ The finite-size scaling plots of the order parameter data $%
M(T_{C}(\infty ),\ell )$ at the infinite lattice critical temperature ($%
\varepsilon =0$ ) are illustrated in figure 4. For $\delta _{\ell }=5$, $%
\beta =0.31$ and $\nu =0.64$ universal values, the order parameter data lie
on a single curve with the slope equal to $1/\delta _{\ell }=0.2$ in the
interval $0.229\leq x\leq 2.875$. The estimated value is in good agreement
with the value of the field critical exponent( $\delta _{h}$=$5$).
Furthermore, the order parameter amplitude in equation (9) is estimated as $%
B_{\ell }=1.096$. This value is in agreement with the linear extrapolation
result (Figure. 2(c)). The finite-size scaling plots of the susceptibility
data $\chi (T_{C}(\infty ),\ell )$ are shown using $\delta _{\ell }=5$, $%
\gamma =1.25$ and $\nu =0.64$ universal values in figure 5. As it is seen in
figure 5, the scaling susceptibility data in the interval $0.236\leq x\leq
3.018$ lie on a straight line with the slope $(1-\delta _{\ell })/\delta
_{\ell }=-0.8$. It gives the value of the dipol-quadrupol critical exponent $%
\delta _{\ell }$ as $5$. The value of the straight line at $\varkappa =0$
gives the susceptibility amplitude as $C_{\ell }=0.063$.

The previous studies implied that the dipol-quadrupol interaction $\ell $ on
the spin-1 Ising model Hamiltonian is considered as a magnetic field-like
perturbation $\left[ 8,35\right] $. Our simulations exposed that the
dipol-quadrupol interaction parameter $\ell $ has prevented the one of the
spin species ($S=+1$ or $-1$) and it has broken the symmetry in the
Hamiltonian. Really, the $\ell $ acts similar to the external magnetic field 
$h$ over the spin system. However, the order parameter $M(T_{C},\ell )$ and
the susceptibility $\chi (T_{C},\ell )$ behave according to the equations of
state ($\ M\thicksim \ell ^{1/\delta }$ and $\chi \thicksim \ell ^{(1-\delta
_{\ell })/\delta _{\ell }}$) in the interval $0\leq \ell =L/J\leq 0.01$ at
the critical temperature. Moreover, the order parameter ($M$) and the
susceptibility ($\chi $) data are scaled well within the framework of the
finite size scaling hypothesis (at $T_{C}$($\infty $) obtained for $\ell =0$%
). The obtained value of the critical exponent $\delta _{\ell }$ are in good
agreement with the universal value ($\delta _{h}=5$) of the external
magnetic field. These results indicate that the dipol-quadrupol interaction
parameter $\ell $ is similar to the external magnetic field $h$ over the
spin system. Therefore, $\ell $ can be considered as a magnetic field-like
perturbation.

\textbf{Acknowledgement}

This work is supported by a grant from Gazi University (BAP:05/2003-07).

\textbf{References}

$\left[ 1\right] $ Blume M, Emery V J and Griffiths R B, 1971 \textit{Phys.
Rev. A 4 1071 }

$\left[ 2\right] $ Ausloos M, Clippe P, Kowalski J M, Pekalski A, 1980 
\textit{Phys. \ Rev. A} \textit{22, 2218}; 1980 \textit{IEEE Trans.
Magnetica MAG 16}

\textit{\ \ \ \ \ \ 233 }

$\left[ 3\right] $ Ausloos M, Clippe P, Kowalski J M, Pekalska J, Pekalski
A, 1983\textit{\ Phys. Rev. A} \textit{28 3080}; Droz M, Ausloos M, Gunton J
D, \textit{ibid.}\ 1\textit{978 18 388 }

$\left[ 4\right] $ Ausloos M, Clippe P, Kowalski J M, Ekalska J P, Pekalski
A, 1983 \textit{J. Magnet. and Magnet. Matter} 39 21

$\left[ 5\right] $ Lajzerowicz J and Siverdi\.{e}re J, 1975 \textit{Phys.
Rev. A 11 2090}

$\left[ 6\right] $ Lajzerowicz J and Siverdi\.{e}re J, 1975 \textit{Phys.
Rev. A 11 2101}

$\left[ 7\right] $ Schick M and Shih W H, 1986 \textit{Phys. Rev. B 34 1797}

$\left[ 8\right] $ Newman K E and Dow J D, 1983 \textit{Phys. Rev. B 27 7495 
}

$\left[ 9\right] $ Gu B L, Newman K E, Fedders P A, 1987 \textit{Phys Rev B
35 9135 }

$\left[ 10\right] $ Gu B L, Ni J, Zhu J L, 1992 \textit{Phys. Rev. B 45 4071 
}

$\left[ 11\right] $ Lawrie I D, Sarbach S, \textit{Phase transitions and
Critical Phenomena, edited by C Domb and J L Lebowitz Vol 9 Academic Press
New York 1984}

$\left[ 12\right] $ Kessler M, Dieterich W and Majhofer A, 2003 \textit{%
Phys. Rev. B \ 67 134201}

$\left[ 13\right] $ Saul D M, Wortis M, Stauffer D, 1974 \textit{Phys. Rev.
B 9 4964 }

$\left[ 14\right] $ Landau D P, Swendsen R H, 1980 \textit{Phys.Rev. Lett.
46 1437 }

$\left[ 15\right] $ Chang T S, Tuthill G F, Stanley H E, 1974 \textit{%
Phys.Rev.B 9 4882 }

$\left[ 16\right] $ Herrmann H J, 1979\textit{\ Z. Phys. B 35 171 }

$\left[ 17\right] $ Lebowitz J L, Gallavotti G, 1971 \textit{J. of Math.
Phys. 12 1129 }

$\left[ 18\right] $ Mukamel D, Blume M, 1974 \textit{Phys. Rev. A 10 610 }

$\left[ 19\right] $ Krinsky S, Furman D, 1975 \textit{Phys.Rev. {}B 11 2602 }

$\left[ 20\right] $ Cai Z X, Mahanti S D, 1987 \textit{Phys. Rev. B 36 6928 }

$\left[ 21\right] $ Lapinskas S, Rosengren A, 1994 \textit{Phys. Rev. B 49
15190}

$\left[ 22\right] $ Collins J B, Rikvold P E, Gawlinski E T, 1988 \textit{%
Phys. Rev. B 38 6741 }

$[23]$ Osorio Roberto, Froyen Sverre, Zunger Alex, 1991 Phys. Rev.B 43 14055

$\left[ 24\right] $ Ekiz C, 2004 \textit{Phys. Lett. A 332 121 }

$\left[ 25\right] $ \"{O}zkan A, Kutlu B, 2010 \textit{Int. J. of Mod. Phys
B, accepted to publish}

$\left[ 26\right] $ Demirel H, \"{O}zkan A, Kutlu B, 2008 \textit{Chin.
phys. Lett. 25 2599 }

$\left[ 27\right] $ \"{O}zkan A, Kutlu B, 2007 \textit{Int. J. of Mod. Phys
C 18 1417 }

$\left[ 28\right] $ \"{O}zkan A, Kutlu B, 2009 \textit{Int. J. of Mod. Phys
C 20 1617 }

$\left[ 29\right] $ \"{O}zkan A, Sefero\u{g}lu N and Kutlu B, 2006 \textit{%
Physica A 362 327 }

$\left[ 30\right] $ Creutz M , 1986 \textit{Ann. Phys.} 167, \textit{62}

$\left[ 31\right] $ Saito K, Takesue S, Miyashita S, 1999 Phys.Rev.E 59 2783

$\left[ 32\right] $ Gwizdalla TM, 2004 Czech. J. Phys. 54 679

$\left[ 33\right] $ Kutlu B, Kasap M and Turan S, 2000 \textit{Int. J. of
Mod. Phys. C 11 561 }

$\left[ 34\right] $ Huang K, \textit{Statistical Mechanics 398 John Wiley \&
Sons,1987}

$\left[ 35\right] $ Berker A N, Wortis M, 1976 \textit{Phys. Rev. B 14 4946}

\textbf{Figure Captions}

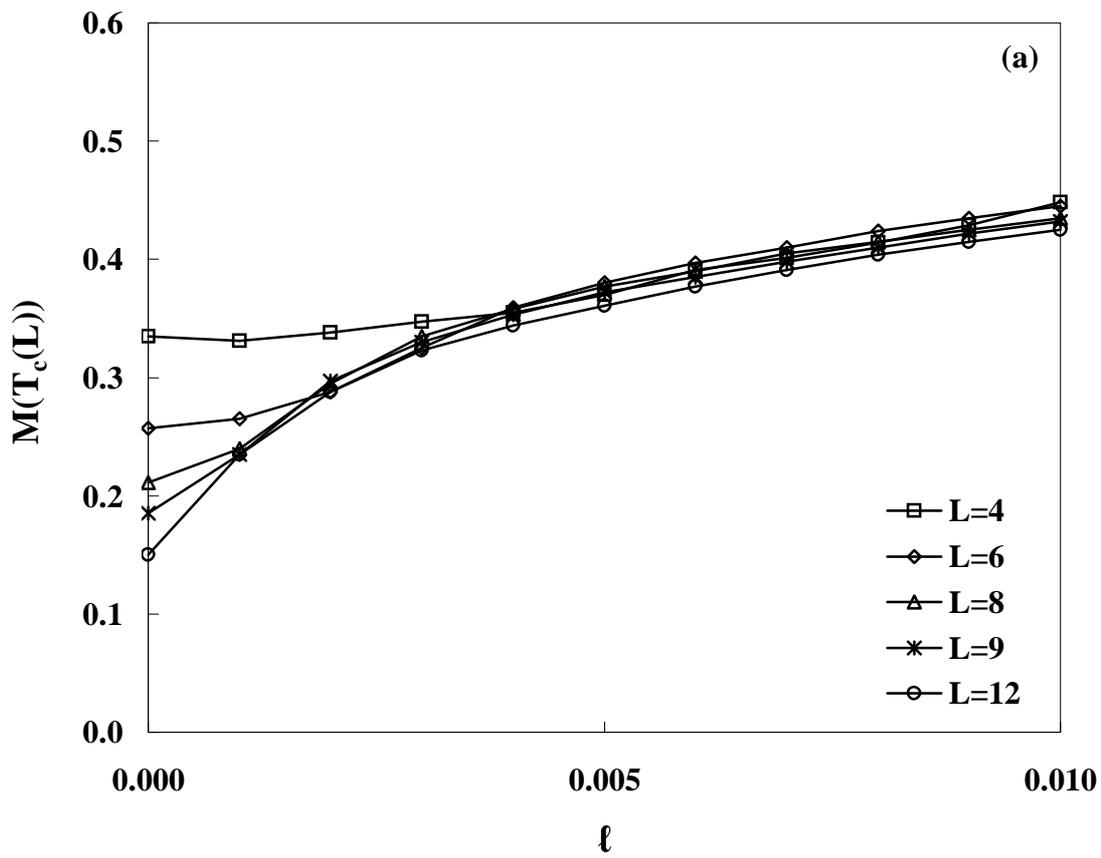
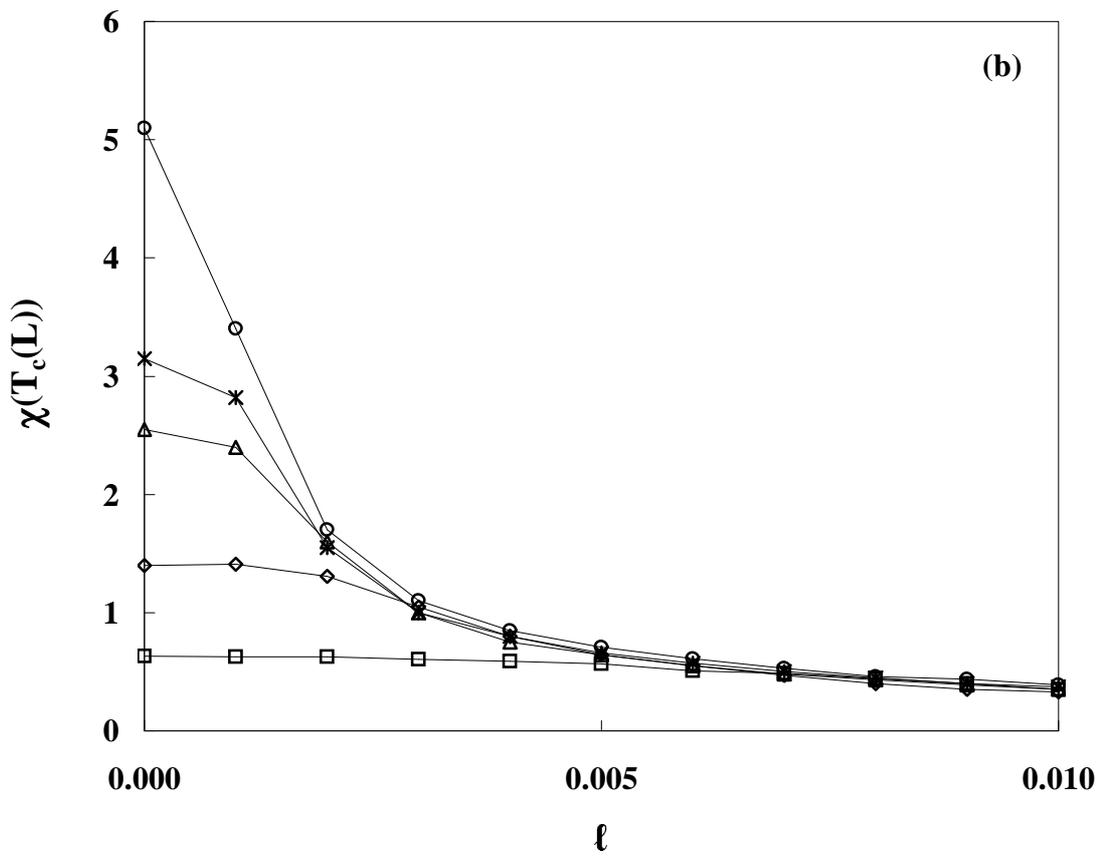
\textbf{Figure.1}. At $T=T_{C}($L$)$ on the different lattice sizes L$=4$, $%
6 $, $8$, $9$ and $12$ in the interval $0\leq \ell \leq 0.001$ for $d=0$ and 
$k=0$ (a) the order parameter values $M$ against $\ell $, the susceptibility
values $\chi $ against $\ell $.

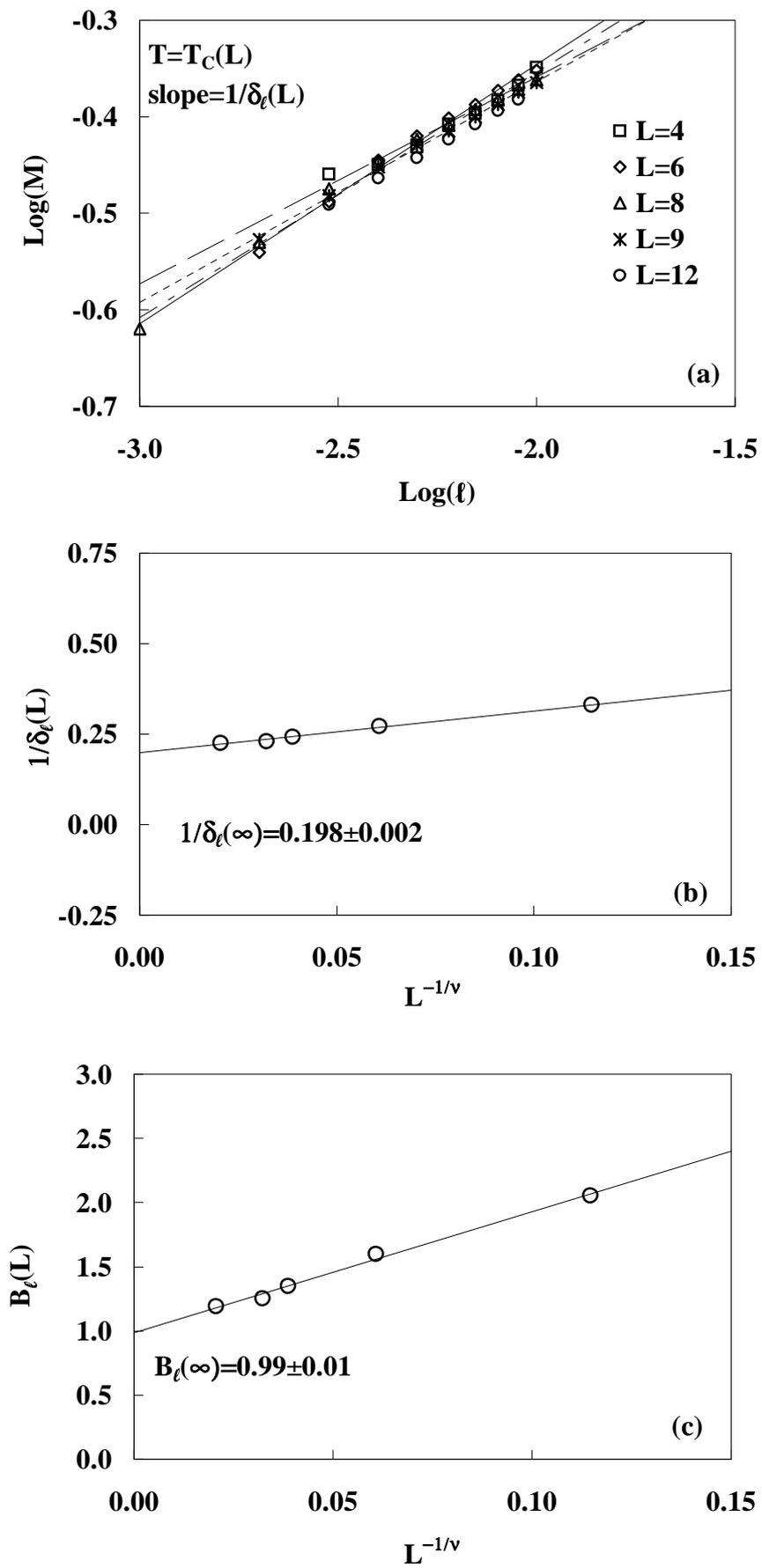
\textbf{Figure.2.} On the different lattice sizes L$=4$, $6$, $8$, $9$ and $%
12$ (a) Log-log plots of $M$ versus $\ell $, (b) The plot of $\delta _{\ell
}($L$)$ versus L$^{-1/\nu }$, (c) The plot of $B_{\ell }($L$)$ versus L$%
^{-1/\nu }$ .

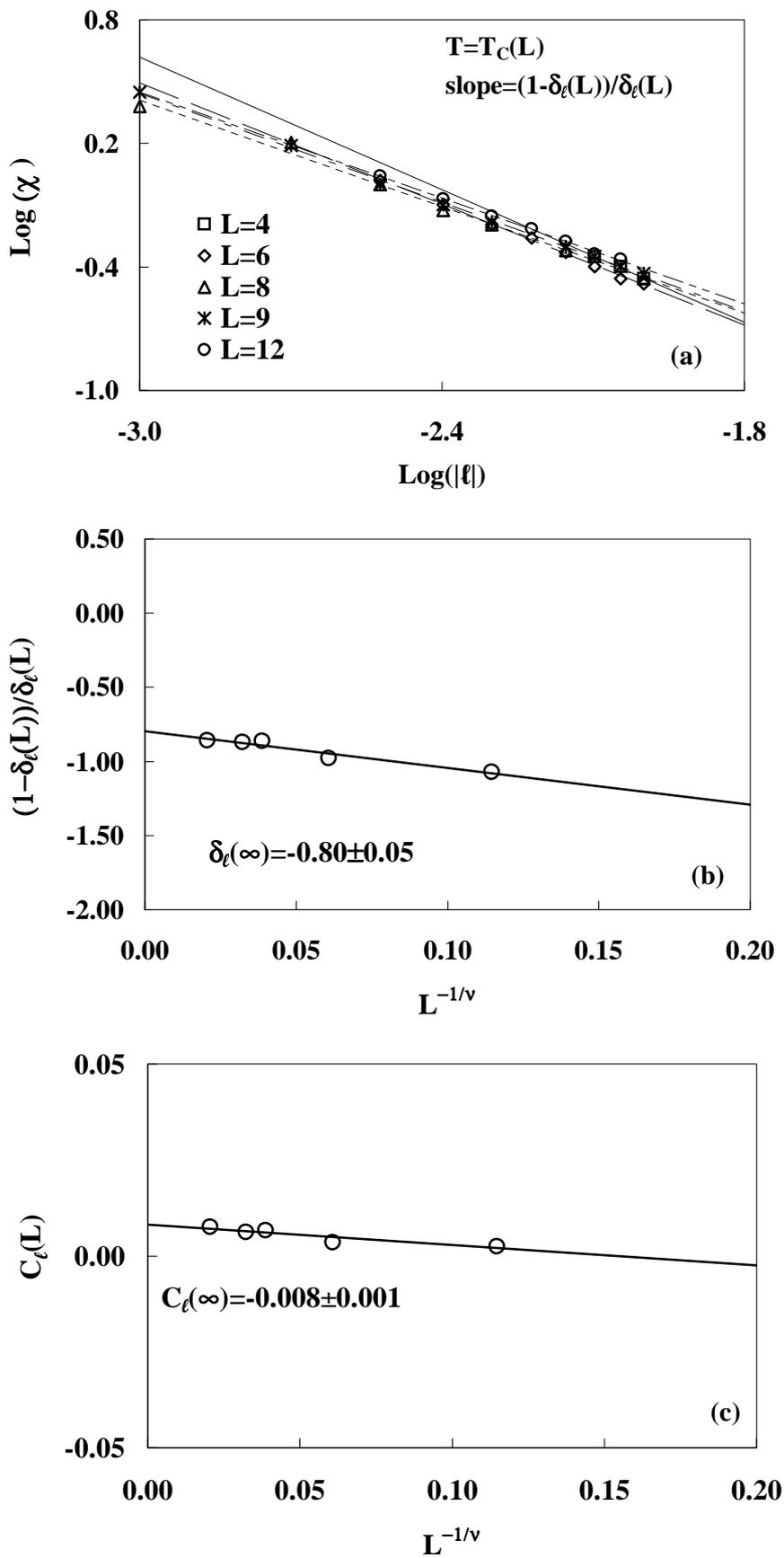
\textbf{Figure.3}. On the different lattice sizes L$=4$, $6$, $8$, $9$ and $%
12$ (a) Log-log plots of $\chi $ versus $\ell $, (b) The plot of $\delta
_{\ell }($L$)$ versus L$^{-1/\nu }$, (c) The plot of $C_{\ell }($L$)$ versus
L$^{-1/\nu }$ .

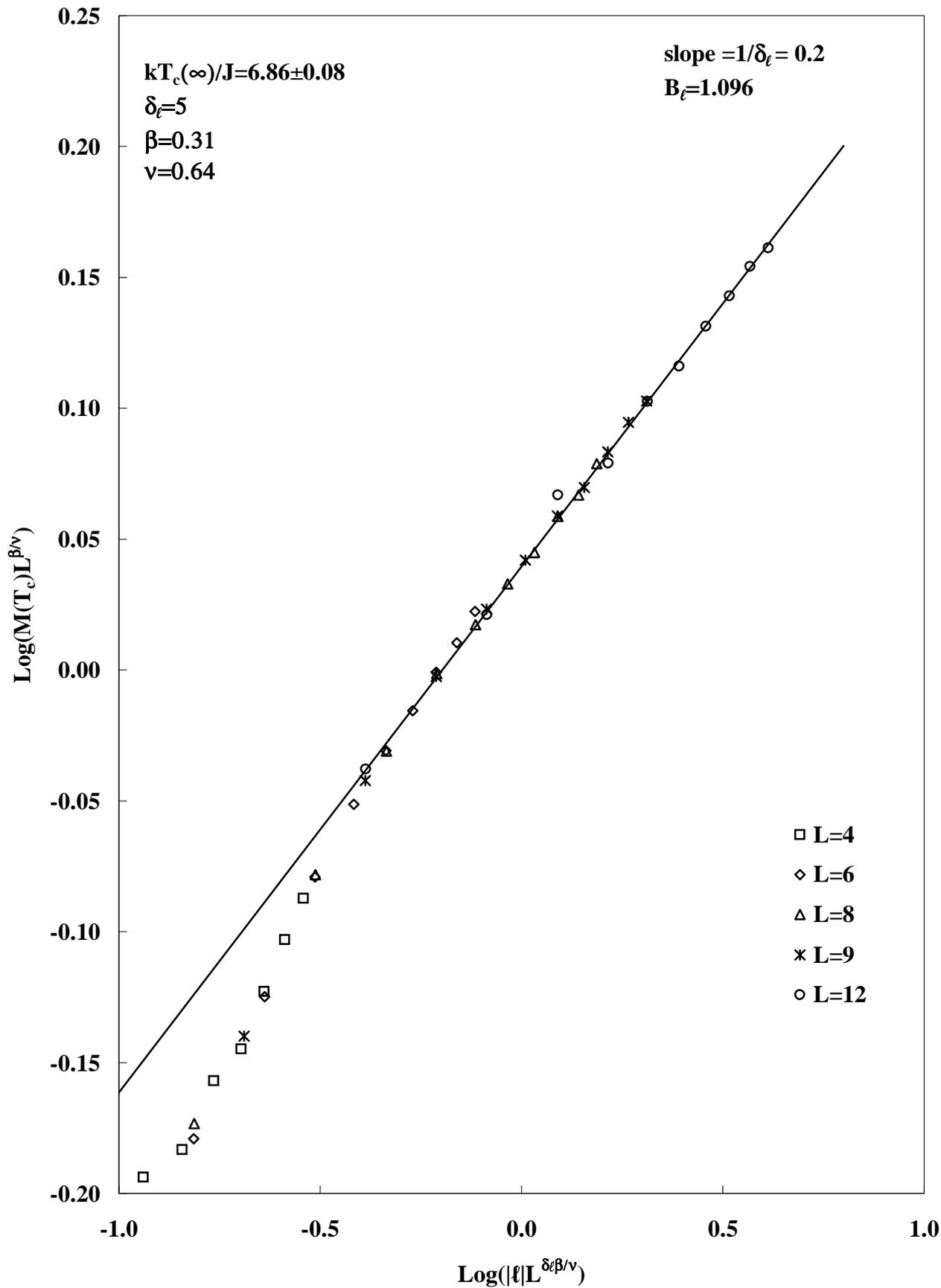
\textbf{Figure.4}. The finite-size scaling plot of the order parameter ($M$)
for $T<T_{C}(\infty )$ on the different lattice sizes L$=4$, $6$, $8$, $9$
and $12$

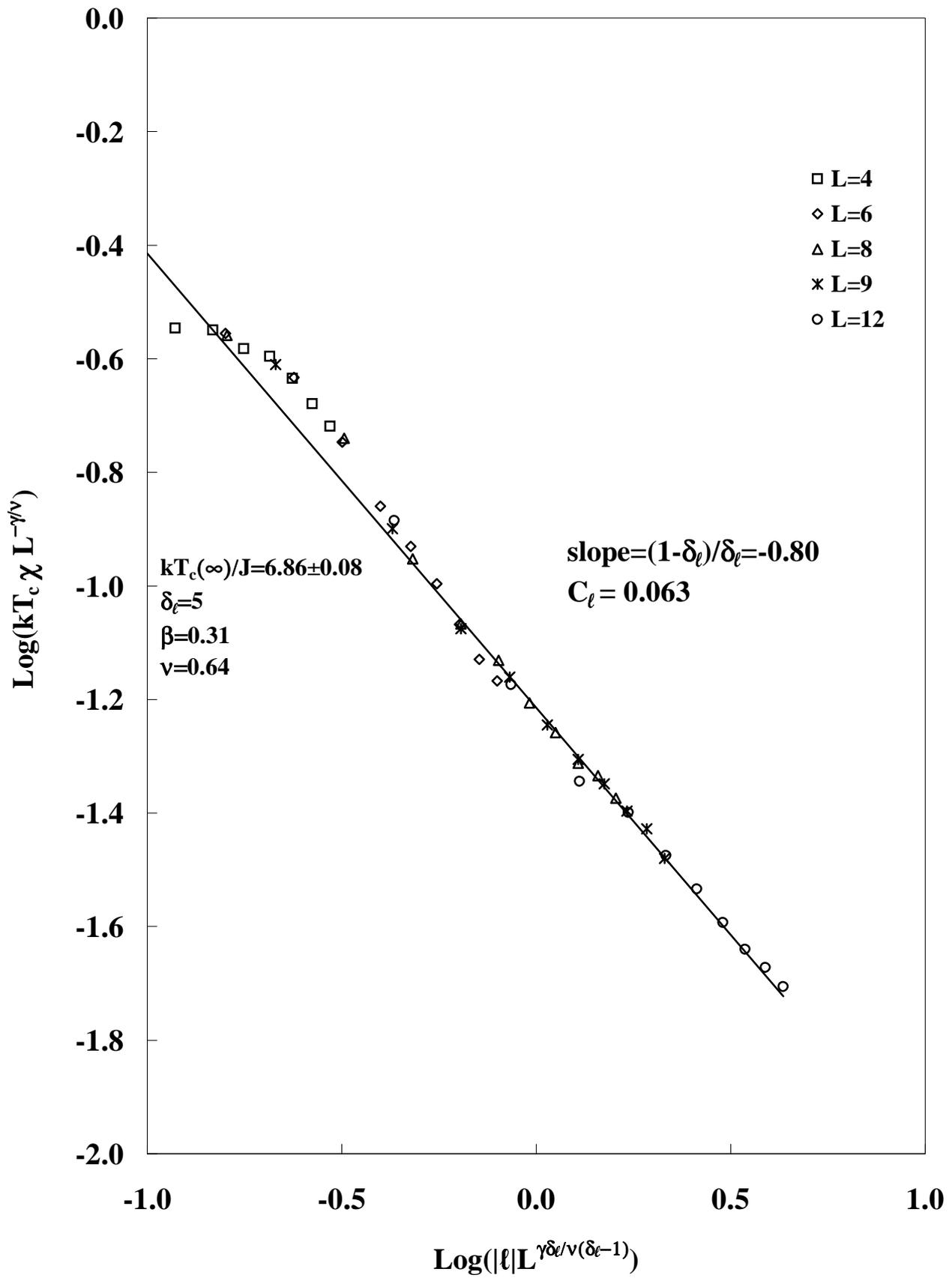
\textbf{Figure 5.} The finite-size scaling plot of the susceptibility ($\chi 
$) for $T<T_{C}(\infty )$ on the different lattice sizes L$=4$, $6$, $8$, $9$
and $12$

\bigskip

\bigskip

\end{document}